\documentclass[a4paper,fleqn,usenatbib]{mnras}

\usepackage{newtxtext,newtxmath}

\usepackage[T1]{fontenc}
\usepackage{ae,aecompl}

\usepackage{graphicx}	
\usepackage{amsmath}	
\usepackage{amssymb}	

\title[Hybrid He-CO white dwarfs]{Formation and evolution of hybrid He-CO white dwarfs and their properties}

\author[Zenati, Toonen \& Perets]{
Yossef Zenati$^1$, Silvia Toonen$^{1,2}$ and Hagai B. Perets$^1$ 
\\
$^1$ Physics Department, Technion - Israel Institute of Technology, Haifa
3200004, Israel\\
$^2$ Anton Pannekoek Institute for Astronomy, University of Amsterdam, 1090 GE, Amsterdam, The Netherlands}

\date{Accepted XXX. Received YYY; in original form ZZZ}

\pubyear{2018}

\begin{document}
\label{firstpage}
\pagerange{\pageref{firstpage}--\pageref{lastpage}}
\maketitle


\begin{abstract}
White dwarfs (WDs) are the stellar core remnants of low mass ($\lesssim8\,{\rm M_{\odot}})$
stars. They are typically divided into three main composition groups:
Oxygen-Neon (ONe), Carbon-Oxygen (CO) and Helium (He) WDs. The evolution
of binary systems can significantly change the evolution of the binary
stellar components. In particular, striping the envelope of an evolved
star can give rise to a core remnant, which can later evolve into
a WD with significantly different composition. Here we focus on the
formation and evolution of hybrid HeCO WDs. We follow the formation
and stellar evolution of such WDs for a range of initial conditions
and provide their detailed structure, mass-radius relation and luminousity-temperature
evolution. We find that both low-mass WDs ($<0.45{\rm M_{\odot}},$
typically thought to be He-WDs) and intermediate-mass WDs ($0.45<{\rm M_{WD}\le0.7},$
typically thought to be CO-WDs) could in fact be hybrid HeCO WDs,
with $5-25\,(75-95)\%$ of their mass in He (CO). We use population
synthesis calculations to infer the birth rate and properties of such
WDs. We find that hybrid HeCO-WD comprise the majority of young ($<2$Gyr)
WDs in binaries, but are more rare among older WDs in binaries. The
high frequency and large He content of such WDs could have an important
role in WD-WD mergers, and may give rise to sub-Chandrasekhar thermonuclear
supernova explosions. 
\end{abstract}



\section{Introduction}

White dwarfs (WDs) are the stellar core remnants of low mass ($\lesssim8\,{\rm M_{\odot}})$
stars that formed following their post main-sequence (MS) evolution.
They are composed mostly of electron-degenerate matter, and are divided
into several types, including WDs composed of carbon-oxygen (CO) and
Oxygen-Neon (ONe) WDs, corresponding to the stellar evolutionary end
points of intermediate and high mass stars, respectively. The minimum
mass of a present day WD formed through stellar evolution of a single
star is in the range ${\rm \sim0.50-0.52M_{\odot}}$ arising from
the lowest mass stars with lifetimes shorter than the age of the universe
($1.01$ down to $0.84{\rm M_{\odot}}$ for ${\rm Z={\rm 1.5Z_{\odot}}}$
down-to ${\rm Z}=0.1{\rm Z_{\odot}}$, respectively, as calculated
from stellar evolution models using the SeBa module (\citealt{Too12}).
Such WDs would be CO WDs. However, binary evolution can change these
outcomes and produce different types of WDs and allowing for a much
lower mass range \citep[e.g. ][and references therein]{Han+02,Reb+11}.

In interacting binaries each of the stellar component may fill its
Roche lobe, and may be stripped of part of its hydrogen/helium-rich
envelope during its evolution on the red giant branch (RGB; or the
asymptotic giant branch, AGB) stage. Such altered evolution can give
rise to qualitatively different evolution and the formation of present
day very low mass (VLM; $<0.45$ ${\rm M_{\odot}}$; \citealp[e.g.][]{Han+02})
WDs. The evolution and final outcomes of the binary evolution strongly
depend on the initial conditions: the mass of the stellar components
and their initial separation. In particular, WDs of masses lower than
0.45 ${\rm M_{\odot}}$ are typically thought to be Helium (He)-WDs
formed through this channel \citep[e.g. ][]{Ibe85,Han+02,Nel+01,Ist+16,Zha+18}.
However, the complex binary evolution channel can give rise to VLM
(as well as more massive)\emph{ hybrid-WDs}, composed of significant
fractions of both CO and He. Such white dwarfs descend from stars
which fill their Roche lobes in the stage of hydrogen burning in a
shell, become hot sub-dwarfs in the He-burning stage, but do not experience
envelope expansion after the formation of a degenerate carbon-oxygen
core \citep{Ibe85,Nel+00}.

The exact definition of a hybrid HeCO WD is somewhat arbitrary; one
can consider a hybrid HeCO WD as any WD in which no less than some
fraction $f$ of the mass is composed of He, and no less of a fraction
$f$ of the mass is composed of CO. In practice, in all of the hybrid
WDs we find the the mass is dominated by CO. Here we focus on such
hybrid WDs; we explore their properties and evolutionary channels,
and discuss their implications on our understanding of WDs and their
structure, which are strongly dependent on the composition. Not less
important, HeCO WDs may have an important role in affecting the outcomes
of WD-mergers, and in particular the possible production of thermonuclear
SNe from double-degenerate WD mergers. For the latter, the existence
of a significant mass in He can catalyze thermonuclear SN explosions
even in sub-Chandrasekhar mass WDs \citep[e.g. ][]{Woo+86,Ibe+87,1990ApJ...361..244L,Bil+07,Wal+11}.
Hence, understanding of hybrid WDs is critical for a wide range of
compact objects as well as their potential explosive mergers.

HeCO WDs have been first discussed by \citet{Ibe85} who suggested
that a significant fraction of the VLM WDs could be HeCO WDs rather
than He (only) WDs. The possibility of HeCO WDs, later termed hybrid
WDs, have been then further discussed (though briefly) in various
contexts \citep{Tut+92,Ibe+97,Nel+00}. Studies by \citet{Alt+04}
and \citet{Pan+07} explored the evolution of VLMs from low mass $(<3.16\,{\rm M_{\odot})}$
progenitor stars, and showed that Oxygen-core VLMs can be formed through
binary evolution, producing WDs in the mass range $0.35-0.45$ ${\rm M_{\odot}}$
mostly composed of Oxygen. Follow-up studies by \citet{Pra+09} had
been able to produce a CO WD of $0.33$ M$_{\odot}$ suggesting it
as the lower limit for such WDs. More recent studies have further
developed the study of VLM He WDs \citep[and references therein]{Ist+16,Zha+18}.
All of these later studies explored the possibility of VLM He WDs,
with little discussion of the potential of hybrid He-CO VLMs. Our
focus is exploring the range of possible hybrid HeCO WDs and their
properties, map the possible range of He to CO mass fractions in these
WDs and characterize their structure and mass-radius relationship.
Our detailed stellar evolution findings (using the MESA code; \citealt{2011ApJS..192....3P,2015ApJS..220...15P})
can be used to direct the more simplified population synthesis models
in order to characterize the general properties of hybrid-WD that
form, their binary system progenitors and the type of double-degenerate
mergers in which they participate. The outcomes of the latter are
further explored in forthcoming publications.

The paper is structured as follows. First (section \ref{sec:MESA-numerics}),
we describe the methods used to explore the detailed stellar evolution
models (using the MESA code) and population studies (using the SeBa
population synthesis code; section \ref{sec:BPS}) for the formation
of hybrid-WDs. We present our main results in section \ref{sec:results}
and then discuss the implications of hybrid-WDs, and summarize (section
\ref{sec:summary}).

\section{Detailed stellar evolution: methods}

\label{sec:MESA-numerics}

Hybrid HeCO WDs can evolve from intermediate mass stars in binary
systems through through a phase of mass transfer through Roch-lobe
overflow (RLOF) or shared common-envelope evolution \citep[see e.g.][]{Iva13}.
In the interacting binary the companion envelope can be stripped following
the formation of a He core and the evolution on the red giant branch
(RGB). The later evolution of the stripped star and the He core is
then significantly altered compared with the uninterrupted evolution
of a non-interacting (single) star. When most of the red giant envelope
is removed, hydrogen shell burning is quelled, the He-core keeps growing,
and the star begins to contract. If the He core is sufficiently massive,
the contraction will eventually trigger He ignition (see \citealt{Ibe85})
and the formation of a CO core; the He to CO ratio will then be determined
by the specific detailed evolution of He burning into CO, and mass-loss
through winds from the envelope.

In order to follow the complex evolution, we begin by considering
a range of initial binary conditions. The initial binary separations
are chosen such that the lower mass binary component will eventually
fill its Roche lobe. The stellar evolutionary tracks of binary components
are followed from the pre-MS stage until the final production of the
WD. We mapped these conditions into the stellar evolution code MESA
and the used the binary mode \citep{2015ApJS..220...15P} to follow
their evolution for the range of initial conditions. We also tested
models where the binary evolution (leading to effective mass-loss)
had been introduced by us into single star models. Mass loss was introduced
in these cases when the conditions for the radius and core mass are
such that RLOF is expected. In practice, this point is not fine-tuned,
and these conditions can be fulfilled over a range of stellar radii
(as indicated in Table \ref{tab:initial-final}) once the star experiences
the first thermal pulse, or in some cases the second or third thermal
pulses. In this range the final outcomes of the evolution produce
qualitatively and quantitatively similar results throughout our simulations,
we therefore show the results only for the choice of the middle point
in the relevant range). 

In all cases where both the binary and single mode models converged
we obtained consistent final outcomes, further validating both these
approaches. In some cases, however, only our altered single mode model
numerically converged, in which case we report the results only from
the single-mode model. The properties of the initial binaries extend
over a range of initial primary component mass ($2.5{\rm M_{\odot}\leq{\rm M_{donor}\leq4{\rm M_{\odot}}}}$)
and mass ratios (${\rm 0.65\lesssim q={\rm M_{donor}/{\rm M_{companion}}}\lesssim0.81}$;
with the exception of one case with $q=0.36$); the detailed models
and their outcomes are summarized in Table \ref{tab:initial-final}.
We also modeled donor stars with lower masses ($<2{\rm M_{\odot}})$,
but these only produced He WDs, or He WDs with very low fraction of
CO; also consistent with results by \citet{Pra+09}. 

Though some models had been difficult to converge numerically, we
were able to resolve the evolution of these binaries through the whole
relevant mass range and mass ratio. In all these cases we were able
to produce a hybrid HeCO WD. In the following we discuss in detail
our assumptions and initial conditions. All of our models (and MESA
inlist files) are openly available at GitHub as to enable simple reproduction
of the results (github.com/Hybrid-WD).

In the following we list our detailed assumptions and considerations. 
\begin{itemize}
\item Metallicity: All models are of of Solar metallicity stars ${\rm Z={\rm Z_{\odot}=0.02}}$. 
\item Mass transfer: we use a range of $10^{-8}-10^{-10}{\rm M_{\odot}yr^{-1}}$
for the $min\ mdot\ for\ implicit$ parameter. 
\item Mixing: We set ${\rm \alpha=l/H_{p}=1.5}$ as the ratio of typical
mixing length to the local pressure scale height, where $l$ is the
mixing length, ${\rm \alpha}$ is a free parameter, and ${\rm H_{p}}$
is the pressure scale height. Semi-convective mixing and thermohaline
mixing are taken to be ${\rm {\rm 0.01}}$ and $2$, respectively
\citep{2016ApJS..227...22F}. 
\item Nuclear reaction network: We use a 75 isotope nuclear reaction network,
containing the relevant isotopes needed for He, carbon and oxygen
burning. 
\item Stopping condition: We stop the evolution once the star becomes fully
degenerate WD and no further evolution, beside WD cooling is observed.
Our condition effectively translates to WD luminosity and temperature
that fall below ${\rm L\leq1.8L_{\odot};\:T_{eff}\leq5.1T_{\odot,eff}}$
(or in some cases, ${\rm L\leq1.12L_{\odot};\:T_{eff}\leq4.94T_{\odot,eff}}$),
respectively. 
\item Mass loss: The effective mass loss is introduced through the use of
Reimers formulation ${\rm \dot{M}=-4\times10^{-13}\eta_{R}\cdot L\cdot R\cdot M/L_{\odot}R_{\odot}M_{\odot}}$
where $\eta$ is in the range ${\rm 0.7\lesssim\eta\lesssim1}$. In
order to ensure that the mass of the star is appropriately adjusted
following the mass loss, we let the star relax afterwards. 
\item Overshoot: The optimal overshoot parameter enabling numerical convergence
for the whole mass range (for ${\rm 2.5-4M_{\odot}}$) was found to
be ${\rm f_{ov}=0.0016}.$ 
\end{itemize}

\section{Population synthesis models}

\label{sec:BPS}

The formation and evolution of interacting binaries producing hybrid-WDs
is simulated with the binary population synthesis (BPS) code \texttt{SeBa}
\citep{Por96,Too12,Too13}. \texttt{SeBa} is a \emph{fast} code for
simulating binary evolution based on parametrized stellar evolution,
including processes such as mass transfer episodes, common-envelope
evolution and stellar winds. We employ \texttt{SeBa} to generate a
large population of binaries on the zero-age MS, we simulate their
subsequent evolution, and extract those that produce hybrid - HeCO
WDs.

It was shown in \citet{Too14} that the main sources of differences
between different BPS codes is due to the choice of input physics
and initial conditions. Here we focus only on a specific set of choices,
and use the detailed stellar evolution models to test the BPS models
predictions for hybrid WDs. A wider range of initial conditions and
physical assumptions will be explored in future work, where hybrid
WDs may play a roles in the evolution of double degenerate (WD-WD,
NS-WD and BH-WD) mergers.

The basic conditions and assumptions we use embody a classical set-up
for BPS calculations, and the binaries are generated and evolved in
the following way: 
\begin{itemize}
\item The primary masses are drawn from a Kroupa IMF \citep{Kro93} with
masses in the range between $0.1-100$ ${\rm M_{\odot}}$. 
\item The secondary masses are drawn from a uniform mass ratio distribution
with $0<q\equiv M_{2}/M_{1}<1$ \citep{Rag10,Duc13,DeR14}. 
\item The orbital separations $a$ follow a uniform distribution in $log(a)$
\citep{Abt83} 
\item The initial eccentricities $e$ follow a thermal distribution \citep{Heg75}. 
\item A binary fraction $\mathcal{B}$ of 50$\%$ which is appropriate for
A-type primaries \citep{Rag10,Duc13}. 
\item Hybrid definition: Unless stated otherwise, in the BPS calculations
a hybrid is defined as a WD with at least 5\% of its mass composed
of He (see Eq.\,\ref{eq:mhe}). 
\item We allow bare degenerate helium cores to ignite if the core gets exposed
(through binary interactions) when its mass is within 0.02${\rm M_{\odot}}$
of the mass where the helium flash happens \citep{Han+02,Nel10}. 
\item We construct two BPS models that differ with respect to the common-envelope
phase. This is a short phase in the evolution of a binary system when
both stars share a common-envelope. Despite its strong effect on the
binary orbit, common-envelope evolution is poorly understood \citep[see e.g.][for a review]{Iva13}.
We replicate model $\alpha\alpha$ and model $\gamma\alpha$ from
\citet{Too17}, where the prior is based on the classical energy balance
during the common-envelope phase, whereas the latter is based on a
balance of angular momentum. 
\end{itemize}
In principle, hybrid HeCO WDs can be formed in two ways, either through
the common envelope or the RLOF channels. In the interacting binary
the hydrogen-rich envelope of the hybrid progenitor can be stripped
following the formation of a He core and the evolution on the RGB
or AGB. The later evolution of the stripped star and the He core is
then significantly altered compared with the uninterrupted evolution
of a non-interacting (single) star. When most of the giant envelope
is removed, hydrogen shell burning is quelled, giving rise to a helium
star, that can be observed as a possible SdB star. If the He core
is sufficiently massive (${\rm M}_{c}>0.32\,{\rm M_{\odot}}$), He
ignition can be triggered (see \citealt{Ibe85}, leading to the formation
of a CO core; the He to CO ratio will then be determined by the specific
detailed evolution. For the intermediate mass stars in this channel,
the He ignites under non-degenerate conditions).

In the second evolutionary channel for lower-mass stars (typically
$<2{\rm M_{\odot}})$, ignition of the He occurs under degenerate
conditions. In this channel the progenitors are stripped close to
the peak of first giant branch, contract, and then ignite \citep[see][for details]{Han+02}.
However, Our detailed stellar evolution models show that WDs produced
through this channel are effectively He WDs which contain only a very
small fraction of CO ($<1\%$ ). Nevertheless, foe completeness we
consider the BPS of such WDs, but divide the contributions between
the main channel and this latter channel, which does not produce bona-fide
hybrid HeCO WDs. As discussed below the contribution from the low-mass
progenitors is small and does not significantly affect any of our
conclusions regarding hybrid WDs. 

\section{Results}

\label{sec:results}

\subsection{Detailed stellar evolution models }

\begin{figure}
\includegraphics[scale=0.3]{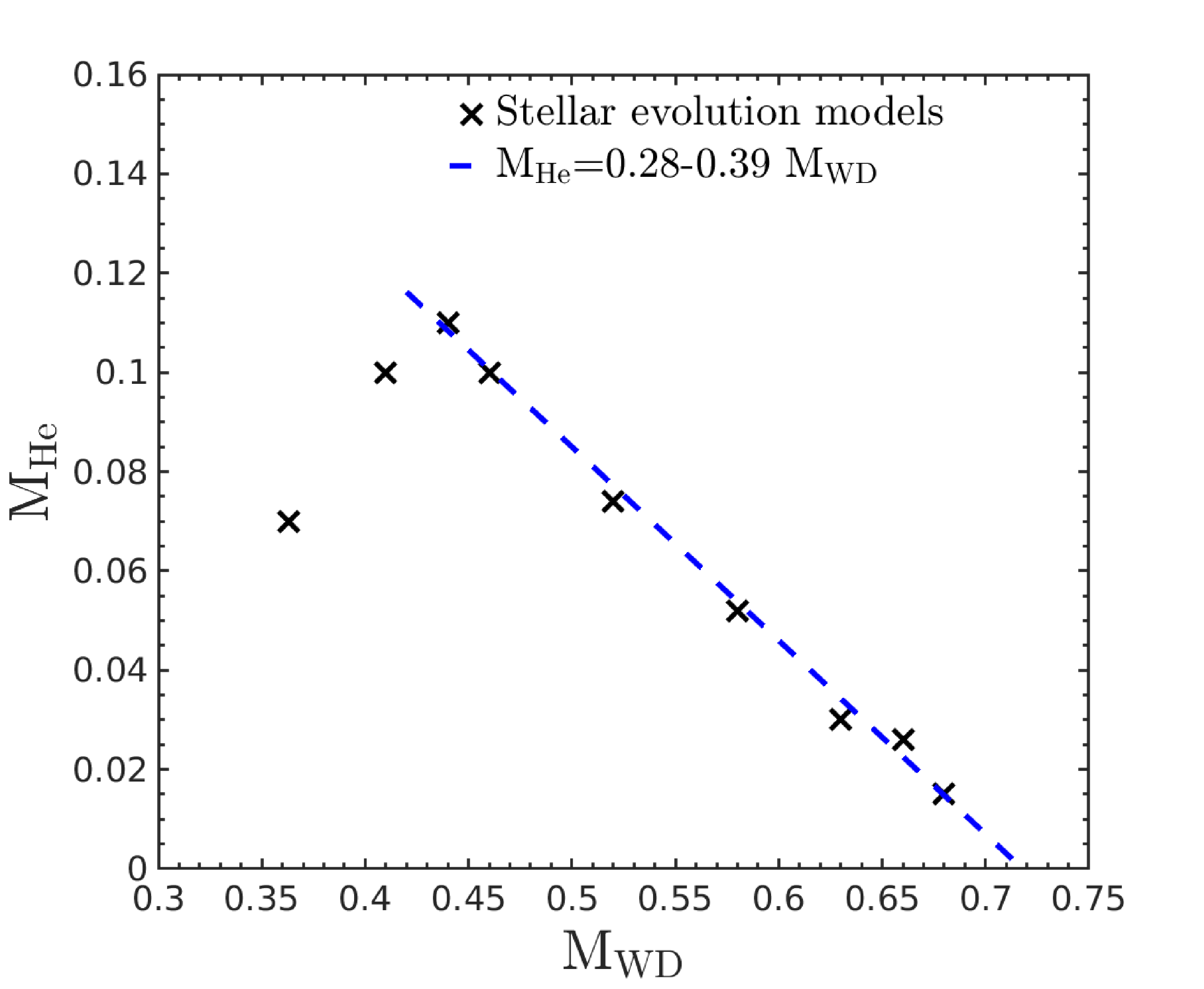}\caption{\label{fig:The-He-mass}The He mass vs. total mass of He WDs from
the detailed stellar evolution models. The blue line depicts a linear
fit for the range $0.4\le{\rm M_{WD}\le0.72}$. }
\end{figure}

\begin{figure*}
\includegraphics[scale=0.5]{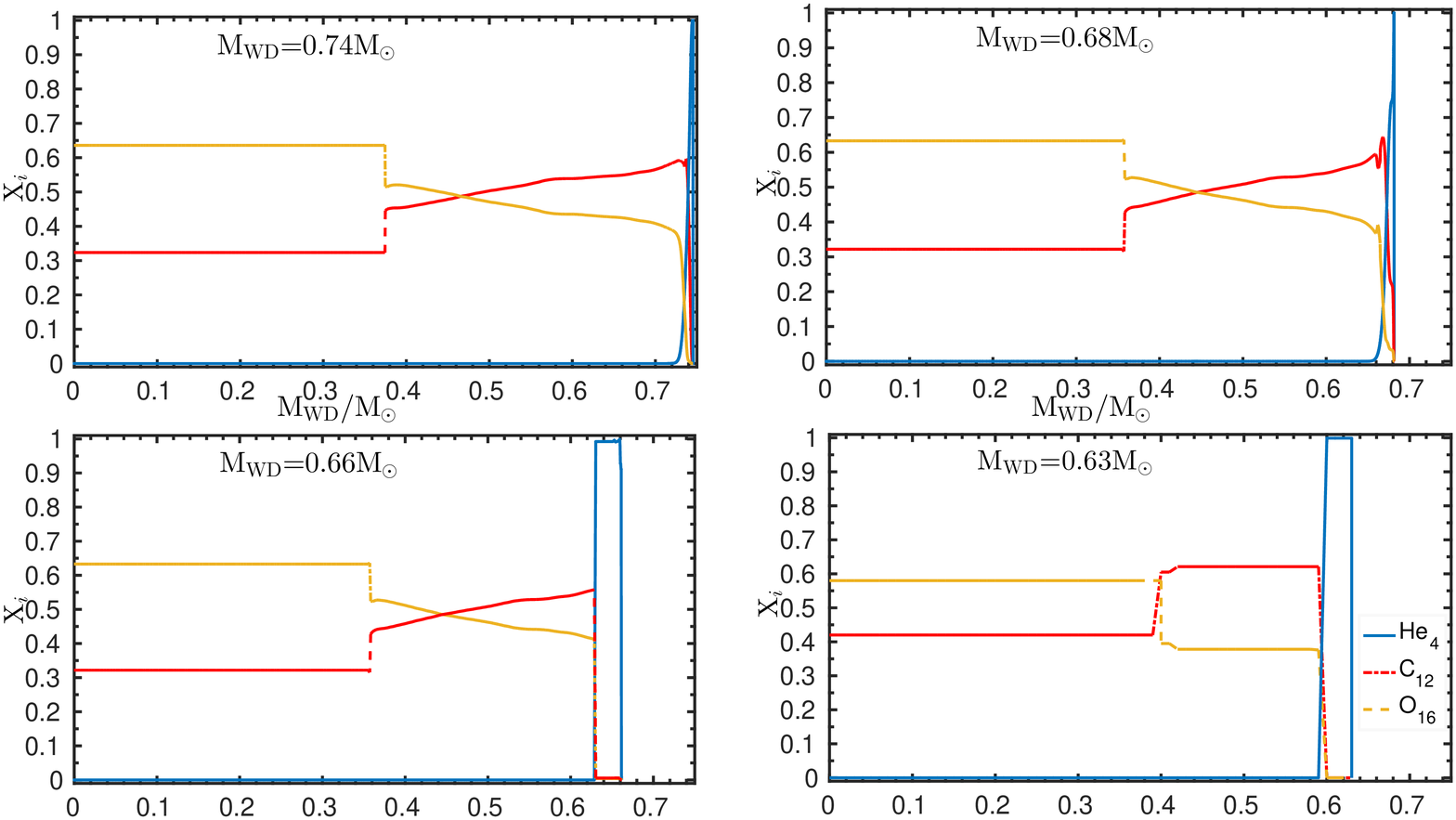}

\includegraphics[scale=0.5]{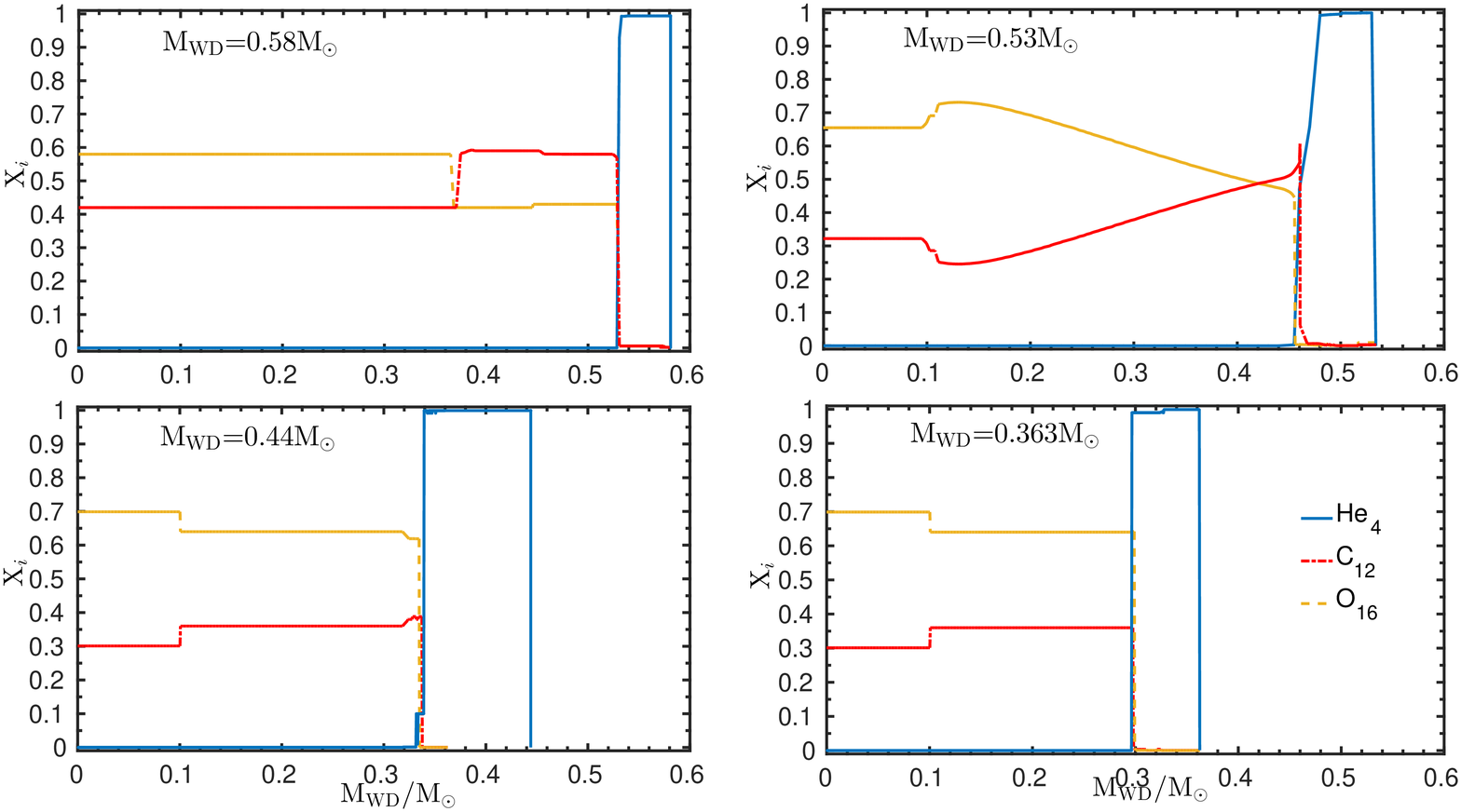}

\caption{\label{fig:composion-structure}The composition structure of hybrid
HeCO WDs for all the calculated models (see Table \ref{tab:initial-final}). }
\end{figure*}

Table \ref{tab:initial-final} and figures \ref{fig:The-He-mass}-\ref{fig:composion-structure}
summarize the results for the mass, composition and structure of hybrid
WDs as a function of their final mass, resulting from our detailed
stellar evolution models. Although hybrids of different masses are
produced from binaries which differ in various aspects (initial mass,
mass ratio, separation etc.), the final compositions form a fairly
continuous sequence for most of the mass range, with He mass fractions
in the range $2-25\%$ (see Fig. \ref{fig:The-He-mass}). As expected,
the He forms an outer shell around the CO core, with only a very small
layer of mixed CO-He composition (see Fig. \ref{fig:composion-structure}).

Our detailed stellar evolution models suggest that the He mass in
the WD $M_{{\rm He}}$ can be approximated by: 
\begin{equation}
{\rm M}_{{\rm He}}=0.28-0.39{\rm M_{WD}},\label{eq:mhe}
\end{equation}
for most of the mass range (${\rm 0.4\le M_{WD}\le0.7}$, where $M_{{\rm WD}}$
is the mass of the WD. This is also approximately consistent with
estimates for the core mass of He stars at which point He burning
is quenched as found by \citet[see their Eq. 89]{Hur00} and implies
there is a maximum mass for hybrids. The maximum mass of a hybrid
WD is therefore $\sim$0.72${\rm M}_{\odot}$, for which the He mass
goes to zero. However, for a more conservative limit, requiring a
significant fraction of the WD mass to be in He (e.g. $>5\%$), the
maximum mass becomes 0.63${\rm M}_{\odot}$. A lower limit to the
mass of a hybrid is expected to be around $\sim$0.32${\rm M}_{\odot}$,
as below the core is not massive enough to ignite He burning \citep[e.g.][]{Han+02}.

\begin{table*}
\begin{tabular}{|c|c|c|c|c|c|c|c|c|c|c|}
\hline 
{\footnotesize{}{}{}{}\#}  & {\scriptsize{}{}{}{}${\rm M_{1}[M_{\odot}]}$}  & {\scriptsize{}{}{}{}${\rm q}$}  & {\scriptsize{}{}{}{}${\rm P[day]}$}  & {\scriptsize{}{}{}{}${\rm logT_{eff}}$}  & {\scriptsize{}{}{}{}${\rm M_{WD}^{f}[M_{\odot}]}$}  & {\scriptsize{}{}{}{}${\rm M_{He4}^{Mesa}[M_{\odot}]}$}  & {\scriptsize{}{}{}{}${\rm M_{C12}^{Mesa}[M_{\odot}]}$}  & {\scriptsize{}{}{}{}${\rm M_{O16}^{Mesa}[M_{\odot}]}$}  & {\scriptsize{}{}{}{}${\rm M_{H1}^{Mesa}[M_{\odot}]}$}  & {\scriptsize{}{}{}{}Modeling$^{*}$}\tabularnewline
\hline 
\hline 
{\footnotesize{}{}{}{}1}  & {\scriptsize{}{}{}{}$2.50$}  & {\scriptsize{}{}{}{}$0.81$}  & {\scriptsize{}{}{}{}$5.01$}  & {\scriptsize{}{}{}{}$5.0$}  & {\scriptsize{}{}{}{}$0.36$}  & {\scriptsize{}{}{}{}$\left(19\%\right),\left(0.07\right)$}  & {\scriptsize{}{}{}{}$\left(42\%\right),\left(0.15\right)$}  & {\scriptsize{}{}{}{}$\left(39\%\right),\left(0.14\right)$}  & {\scriptsize{}{}{}{}$5.4\times10^{-4}$}  & {\scriptsize{}{}{}{}S,B}\tabularnewline
\hline 
{\footnotesize{}{}{}{}2}  & {\scriptsize{}{}{}{}$2.66$}  & {\scriptsize{}{}{}{}$0.78$}  & {\scriptsize{}{}{}{}$4.5$}  & {\scriptsize{}{}{}{}$5.0$}  & {\scriptsize{}{}{}{}$0.41$}  & {\scriptsize{}{}{}{}$\left(25\%\right),\left(0.1\right)$}  & {\scriptsize{}{}{}{}$\left(38\%\right),\left(0.16\right)$}  & {\scriptsize{}{}{}{}$\left(37\%\right),\left(0.15\right)$}  & {\scriptsize{}{}{}{}$5.0\times10^{-4}$}  & {\scriptsize{}{}{}{}B}\tabularnewline
\hline 
{\footnotesize{}{}{}{}3}  & {\scriptsize{}{}{}{}$2.75$}  & {\scriptsize{}{}{}{}$0.79$}  & {\scriptsize{}{}{}{}$5.00$}  & {\scriptsize{}{}{}{}$5.0$}  & {\scriptsize{}{}{}{}$0.44$}  & {\scriptsize{}{}{}{}$\left(25\%\right),\left(0.11\right)$}  & {\scriptsize{}{}{}{}$\left(37\%\right),\left(0.16\right)$}  & {\scriptsize{}{}{}{}$\left(38\%\right),\left(0.17\right)$}  & {\scriptsize{}{}{}{}$6\times10^{-4}$}  & {\scriptsize{}{}{}{}S}\tabularnewline
\hline 
{\footnotesize{}{}{}{}4}  & {\scriptsize{}{}{}{}$2.85$}  & {\scriptsize{}{}{}{}$0.36$}  & {\scriptsize{}{}{}{}$5.25$}  & {\scriptsize{}{}{}{}$5.2$}  & {\scriptsize{}{}{}{}$0.46$}  & {\scriptsize{}{}{}{}$\left(21\%\right),\left(0.10\right)$}  & {\scriptsize{}{}{}{}$\left(38\%\right),\left(0.17\right)$}  & {\scriptsize{}{}{}{}$\left(41\%\right),\left(0.19\right)$}  & {\scriptsize{}{}{}{}$3\times10^{-4}$}  & {\scriptsize{}{}{}{}B}\tabularnewline
\hline 
{\footnotesize{}{}{}{}5}  & {\scriptsize{}{}{}{}$3.01$}  & {\scriptsize{}{}{}{}$0.74$}  & {\scriptsize{}{}{}{}$6.28$}  & {\scriptsize{}{}{}{}$5.0$}  & {\scriptsize{}{}{}{}$0.53$}  & {\scriptsize{}{}{}{}$\left(14\%\right),\left(0.074\right)$}  & {\scriptsize{}{}{}{}$\left(43\%\right),\left(0.23\right)$}  & {\scriptsize{}{}{}{}$\left(43\%\right),\left(0.23\right)$}  & {\scriptsize{}{}{}{}$2\times10^{-5}$}  & {\scriptsize{}{}{}{}S,B}\tabularnewline
\hline 
{\footnotesize{}{}{}{}}6  & {\scriptsize{}{}{}{}$3.20$}  & {\scriptsize{}{}{}{}$0.71$}  & {\scriptsize{}{}{}{}$6.28$}  & {\scriptsize{}{}{}{}$5.2$}  & {\scriptsize{}{}{}{}$0.58$}  & {\scriptsize{}{}{}{}$\left(9\%\right),\left(0.052\right)$}  & {\scriptsize{}{}{}{}$\left(45\%\right),\left(0.26\right)$}  & {\scriptsize{}{}{}{}$\left(46\%\right),\left(0.26\right)$}  & {\scriptsize{}{}{}{}$4.1\times10^{-5}$}  & {\scriptsize{}{}{}{}S,B}\tabularnewline
\hline 
{\footnotesize{}{}{}{}7}  & {\scriptsize{}{}{}{}$3.41$}  & {\scriptsize{}{}{}{}$0.70$}  & {\scriptsize{}{}{}{}$7.00$}  & {\scriptsize{}{}{}{}$5.2$}  & {\scriptsize{}{}{}{}$0.63$}  & {\scriptsize{}{}{}{}$\left(5\%\right),\left(0.03\right)$}  & {\scriptsize{}{}{}{}$\left(48\%\right),\left(0.3\right)$}  & {\scriptsize{}{}{}{}$\left(47\%\right),\left(0.29\right)$}  & {\scriptsize{}{}{}{}$3.0\times10^{-4}$}  & {\scriptsize{}{}{}{}S}\tabularnewline
\hline 
{\footnotesize{}{}{}{}8}  & {\scriptsize{}{}{}{}$3.50$}  & {\scriptsize{}{}{}{}$0.77$}  & {\scriptsize{}{}{}{}$7.44$}  & {\scriptsize{}{}{}{}$5.2$}  & {\scriptsize{}{}{}{}$0.66$}  & {\scriptsize{}{}{}{}$\left(4\%\right),\left(0.026\right)$}  & {\scriptsize{}{}{}{}$\left(48\%\right),\left(0.31\right)$}  & {\scriptsize{}{}{}{}$\left(48\%\right),\left(0.31\right)$}  & {\scriptsize{}{}{}{}$3.2\times10^{-4}$}  & {\scriptsize{}{}{}{}S,B}\tabularnewline
\hline 
{\footnotesize{}{}{}{}}9  & {\scriptsize{}{}{}{}$3.72$}  & {\scriptsize{}{}{}{}$0.66$}  & {\scriptsize{}{}{}{}$8.50$}  & {\scriptsize{}{}{}{}$5.2$}  & {\scriptsize{}{}{}{}$0.68$}  & {\scriptsize{}{}{}{}$\left(2\%\right),\left(0.015\right)$}  & {\scriptsize{}{}{}{}$\left(49\%\right),\left(0.33\right)$}  & {\scriptsize{}{}{}{}$\left(49\%\right),(0.33)$}  & {\scriptsize{}{}{}{}$1.6\times10^{-4}$}  & {\scriptsize{}{}{}{}S,B}\tabularnewline
\hline 
{\footnotesize{}{}{}{}10}  & {\scriptsize{}{}{}{}$4.00$}  & {\scriptsize{}{}{}{}$0.78$}  & {\scriptsize{}{}{}{}$11.13$}  & {\scriptsize{}{}{}{}$5.2$}  & {\scriptsize{}{}{}{}$0.74$}  & {\scriptsize{}{}{}{}$\left(<1.5\%\right),\left(<0.01\right)$}  & {\scriptsize{}{}{}{}$\left(49\%\right),\left(0.36\right)$}  & {\scriptsize{}{}{}{}$\left(49\%\right),\left(0.36\right)$}  & {\scriptsize{}{}{}{}$7.3\times10^{-4}$}  & {\scriptsize{}{}{}{}S,B}\tabularnewline
\hline 
\end{tabular}

$^{*}$Type of modeling using the binary (B) evolution mode in MESA
or an effective binary evolution using a single (S) star module. \caption{\label{tab:initial-final}Characteristics of a binary succeed to survive
until the hybrid WD phase. The ${\rm M_{1}}$ is the donor mass, ${\rm q}$
is the mass ratio ,${\rm P}$ is the initial period for the binary
in RLOF. ${\rm logT_{eff}}$ is the effective temperature when the
donor get our condition. ${\rm M_{WD}^{f}}$ is the final mass of
the donor (hybrid WD mass), ${\rm M_{He4}^{f},M_{C12}^{f},M_{O16}^{f}}$
is the final amount of ${\rm {\rm He4,C12,O16}}$}
\end{table*}

\subsubsection{Evolution on the HR diagram}

\begin{figure}
\includegraphics[scale=0.45]{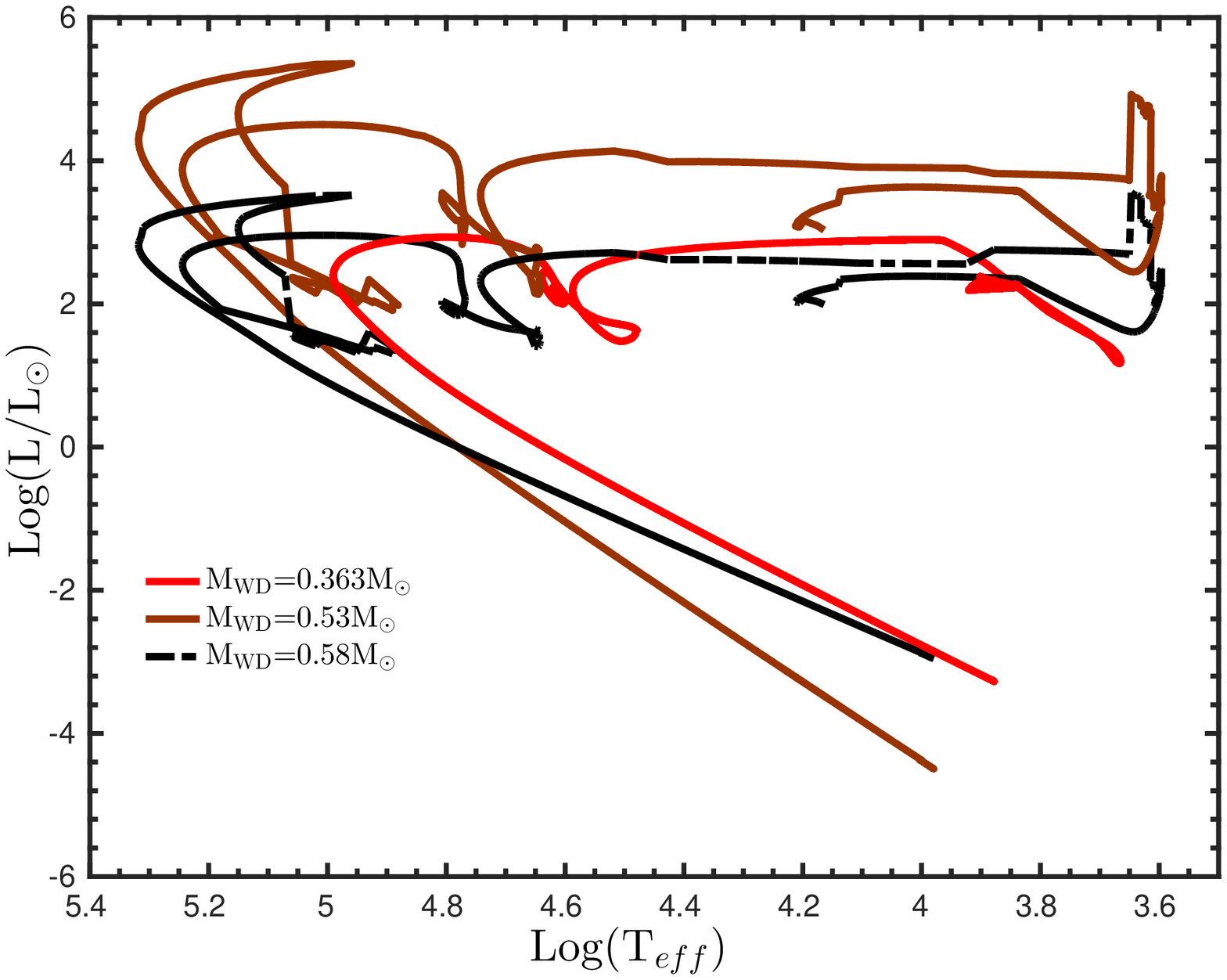}

\includegraphics[scale=0.45]{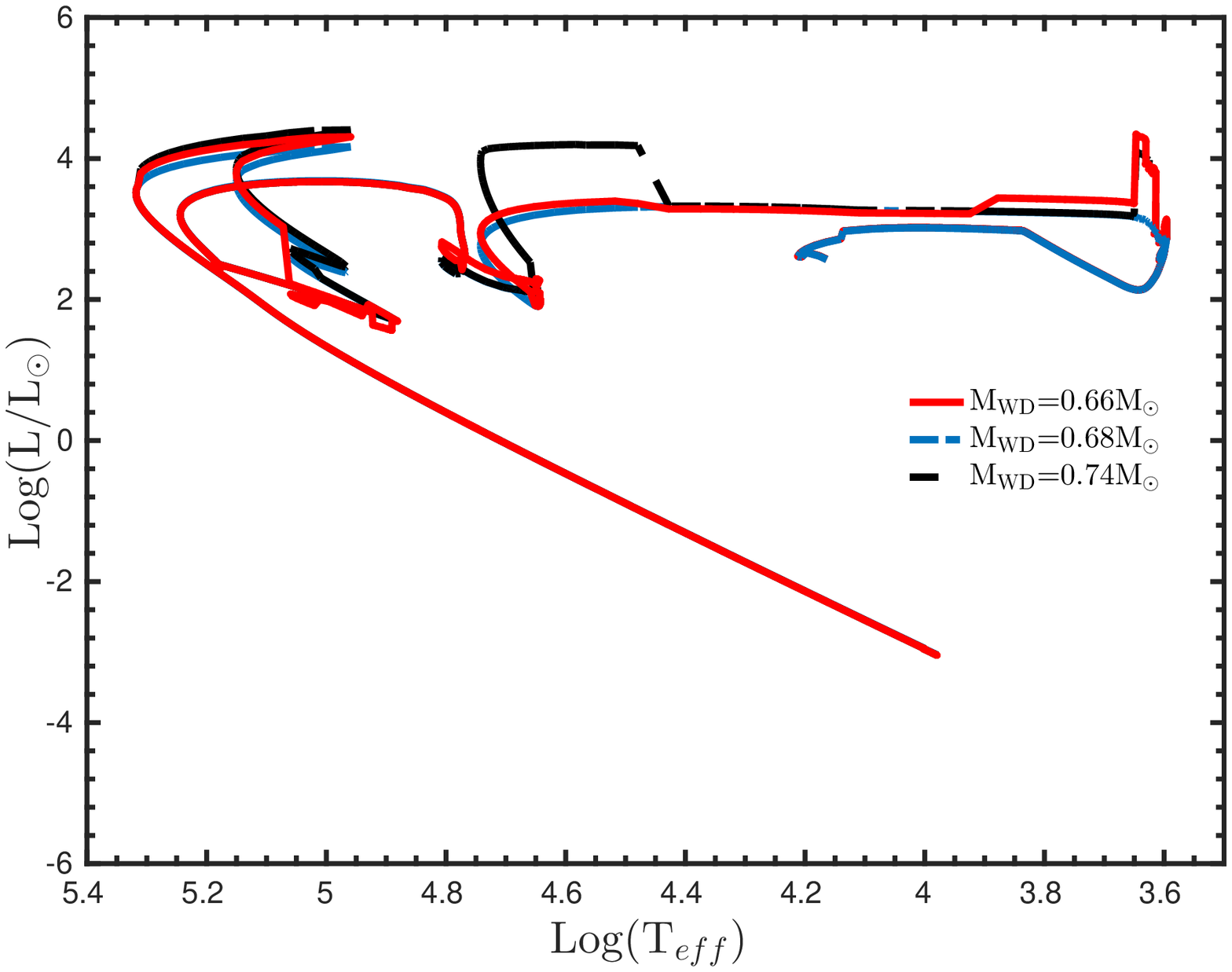}

\caption{\label{fig:HR_all}The HR diagram for the evolution of binaries producing
hybrid HeCO WDs. }
\end{figure}

\begin{figure}
\includegraphics[scale=0.45]{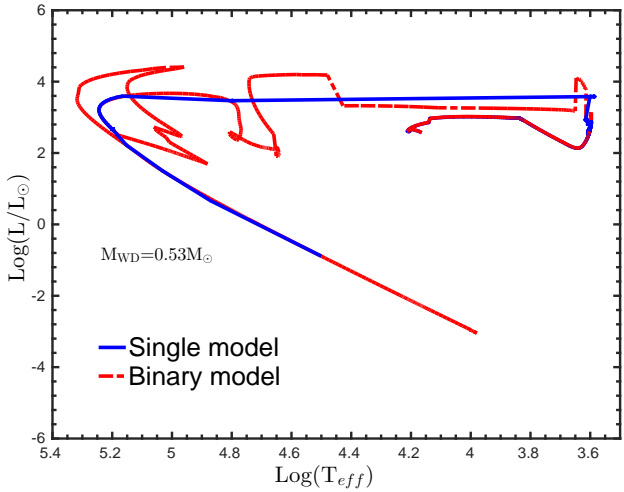}

\caption{\label{fig:binary-vs-single}Comparison of the evolution using the
single and binary modes. Both models begin at the same position, and
mass stripping is modeled as to occur at the same evolutionary stage
in both models (specifically we stripping begins once ${\rm L_{nuc}/L_{ZAMS}}>0.901$
in this case). Though the early evolution is naturally different,
the two evolutionary modes converge once the effective mass loss used
effectively mimic the binary evolution. The excellent correspondence
between the models supports our use of single mode models for the
cases where binary evolution mode did not numerically converge. }
\end{figure}

In Fig. \ref{fig:HR_all} we show the evolution of the HeCO WD progenitor
on the HR diagram for six of the models in Table \ref{tab:initial-final}
covering the full mass range (the other three are not shown for clarity,
but they behave very similarly). As can be seen the evolution can
be quite complex, and though we are mostly interested in the final
WD configuration we refer the interested reader to \citet{Ibe85}
for a detailed discussion of the various evolutionary stages before
the final formation of the hybrid WD. In Fig. \ref{fig:binary-vs-single}
we compare the results from the MESA binary evolution model to a more
simplified single star evolution where the envelope stripping is included
artificially without fully following the binary evolution. As can be
seen the single models follow a different evolution at early times,
but following the stripping they results in the same WD configuration
and evolution, supporting their general use for modeling hybrids.
This also suggests that uncertainties in the mass-transfer phase do
not significantly affect the final structure of the hybrid WDs. Overall
we find only small differences in the final WD configurations when
we use the single and binary models, with typical C, O and He masses
differences of less than 1$\%$ (besides the case of ${\rm M_{WD}}=0.58$
where a $5\%$ difference in the He abundance was found).

\subsection{Structure and composition of of HeCO hybrid WDs}

\begin{figure}
\includegraphics[scale=0.25]{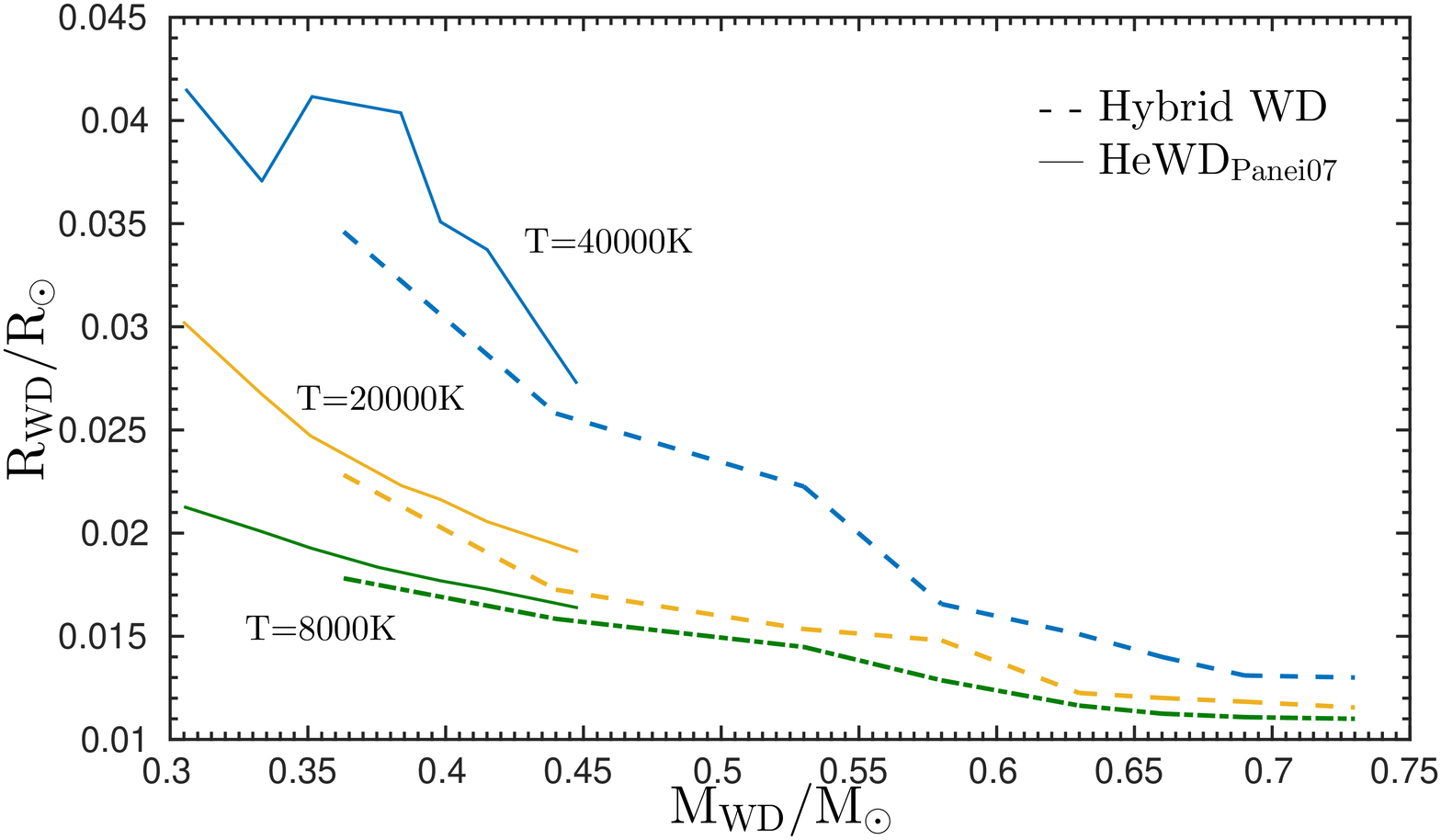}

\includegraphics[scale=0.25]{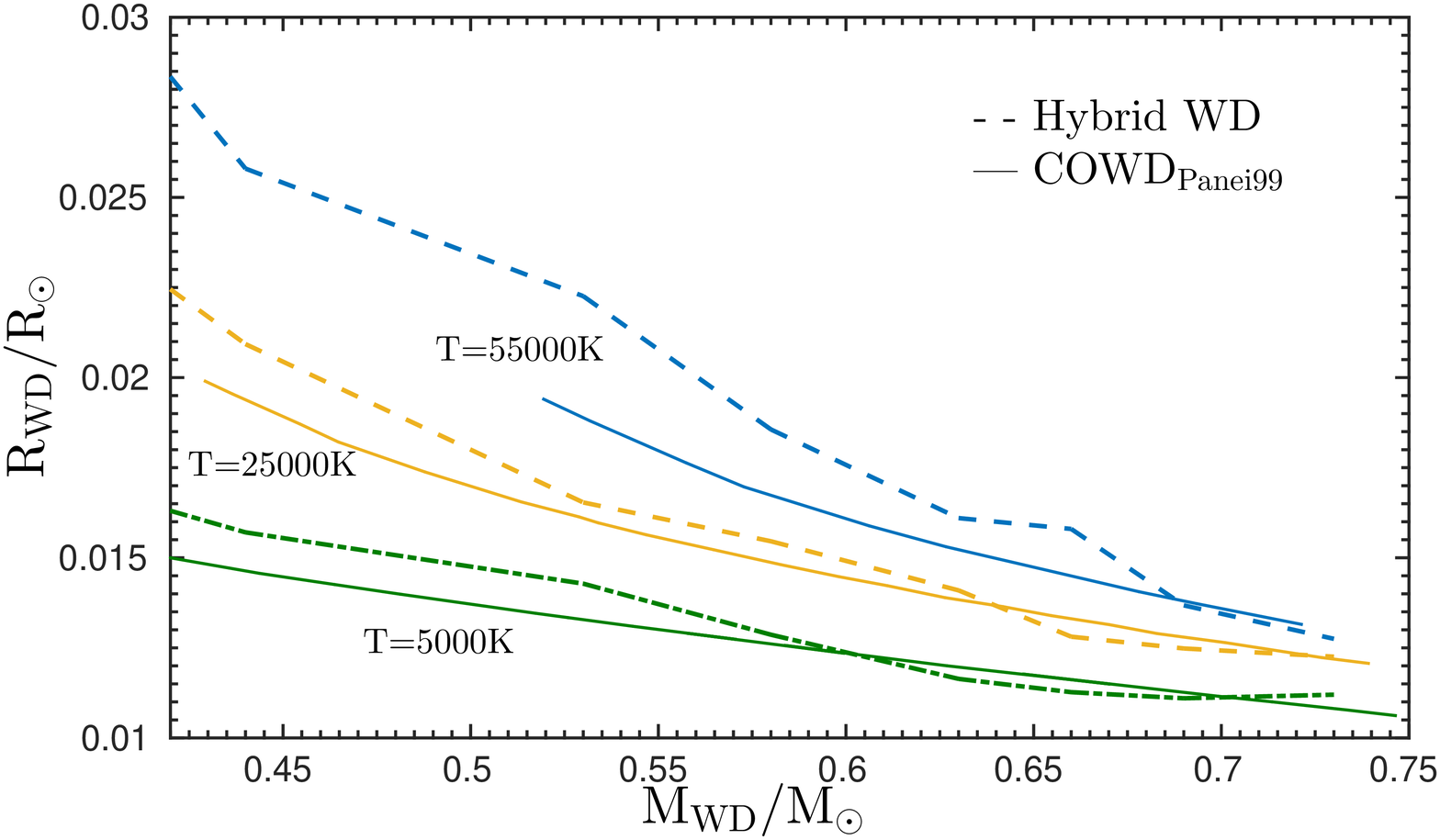}
\caption{\label{fig:mass-radius}The mass radius relation for hybrid HeCO WDs.
The mass radius relation is shown for three temperatures for each
regime (comparison with He WDs on top and with CO WDs in bottom figure).
As expected the radii of Hybrid WDs falls in an intermediate regime
between purely He WDs purely~CO WDs (e.g. compare with \citealt{Pan+00,Pan+07}),
besides for the most massive hybrids where the He fraction is small
and the radii are comparable with that of CO WDs with a small He envelope
(taken from \citealt{Pan+00,Pan+07}).}
\end{figure}

The mass-radius relation for the HeCO WDs is shown in Fig. \ref{fig:mass-radius}.
As expected the HeCO WDs radii typically fall in between those of
CO WDs and purely He WDs.

\subsection{Population synthesis results}

\begin{figure*}
\includegraphics[scale=0.45]{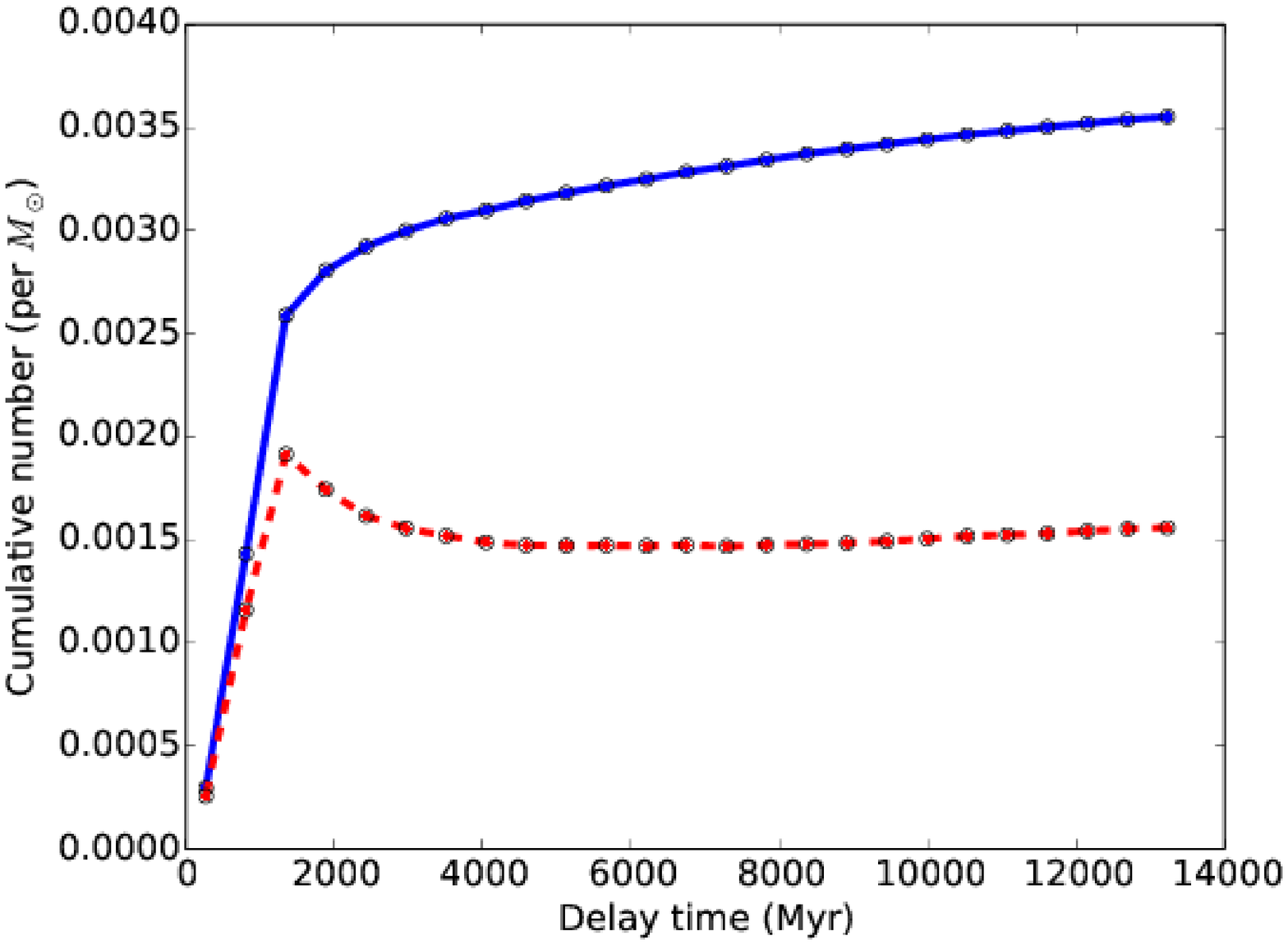}\includegraphics[scale=0.45]{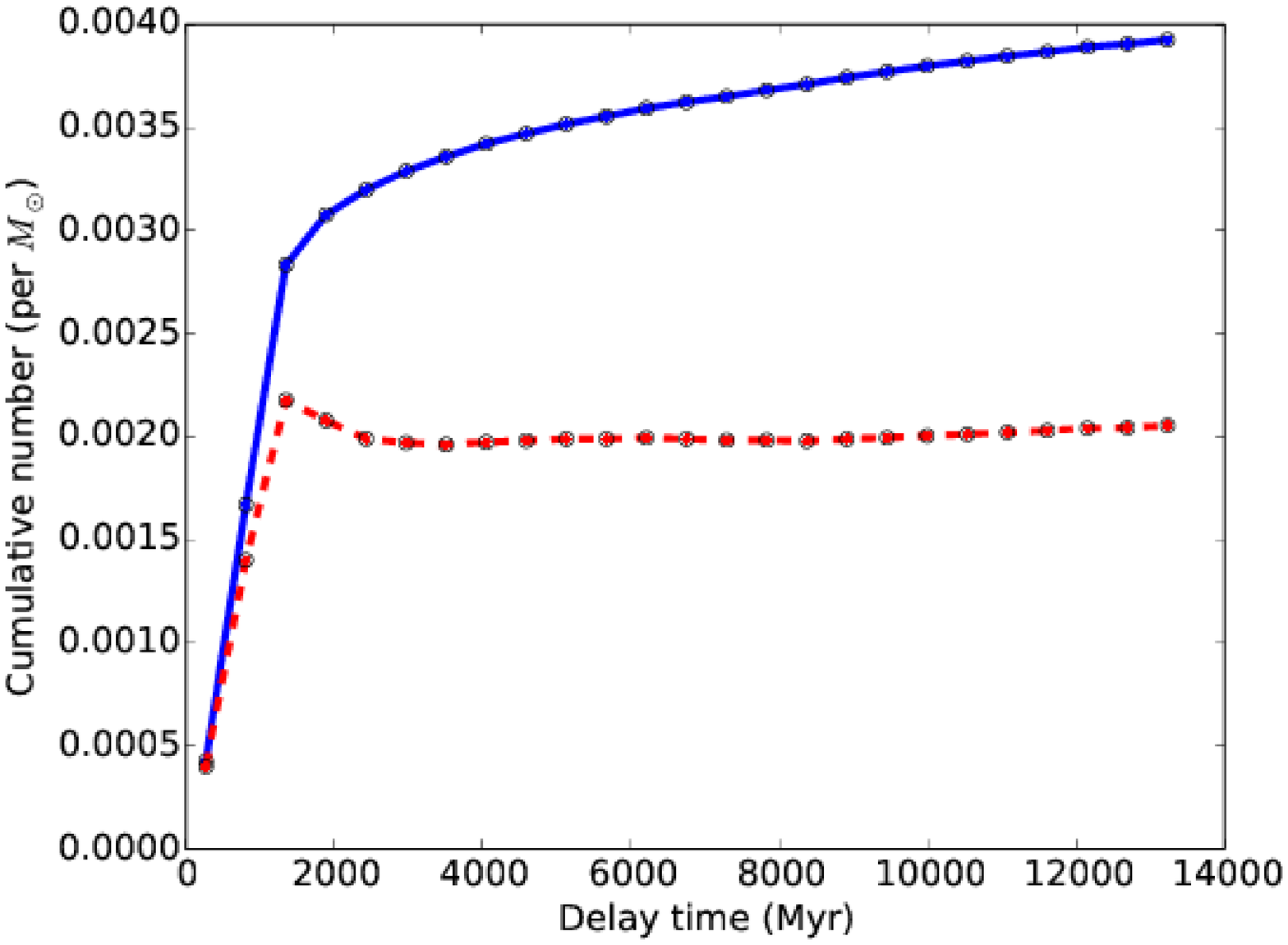}

\caption{\label{fig:lifetime}The total cumulative number (per solar mass)
of hybrid-WD formed as a function of time. On the left model $\alpha\alpha$
for the CE-phase is shown, on the right model $\gamma\alpha$. Upper
(blue) lines show the time integrated birth-rate of hybrid-WDs, i.e.
the cumulative number of \emph{formed} hybrid-WDs as a function of
the delay time. Lower (red) line show the total cumulative number
of \emph{existing} hybrids as a function of the delay time, i.e. after
accounting for the actual lifetime - subtracting the number of hybrids
destroyed (typically through mergers with their companion) from the
total number of formed hybrids.}
\end{figure*}

\begin{figure*}
\includegraphics[scale=0.45]{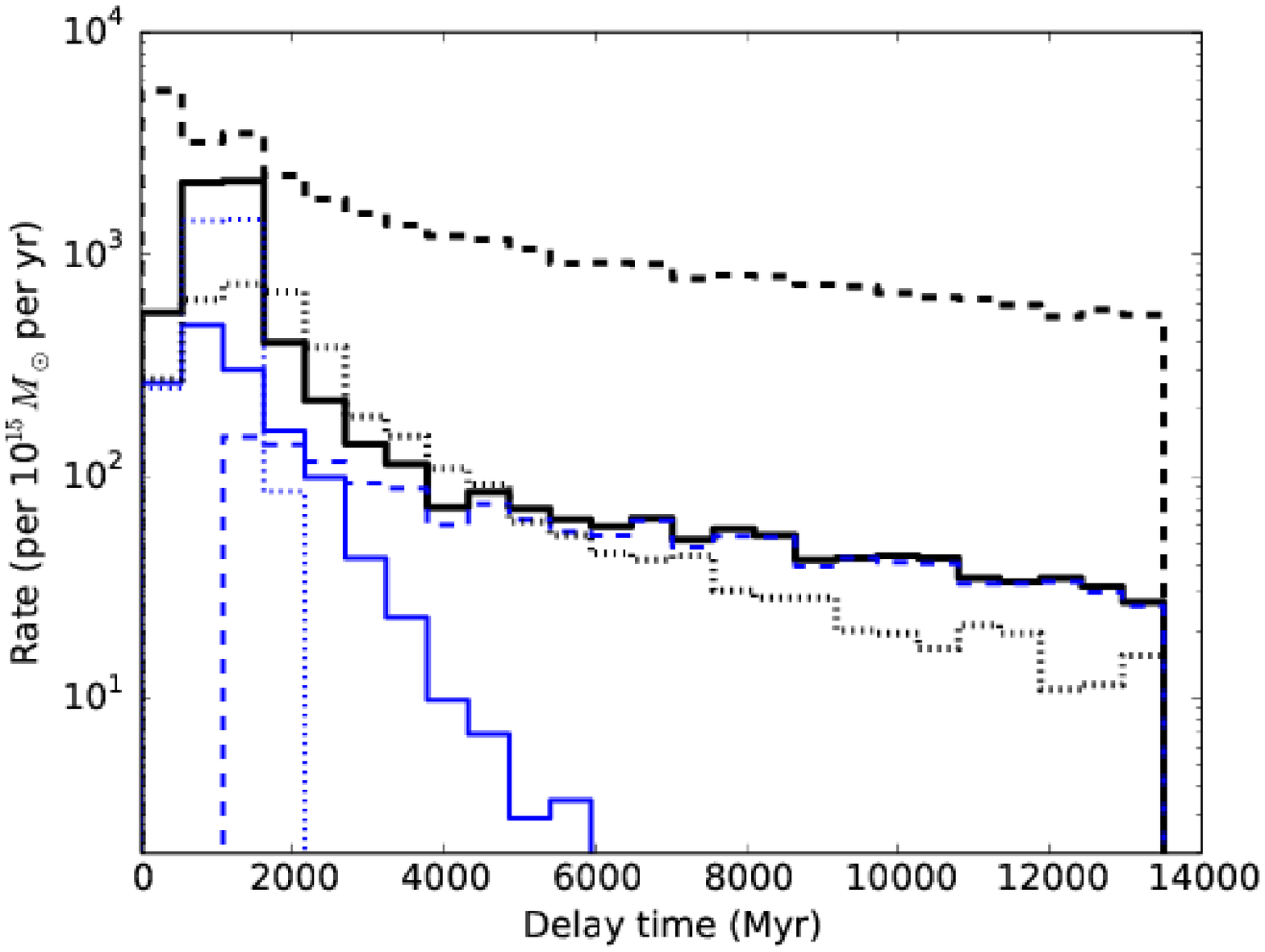}\includegraphics[scale=0.45]{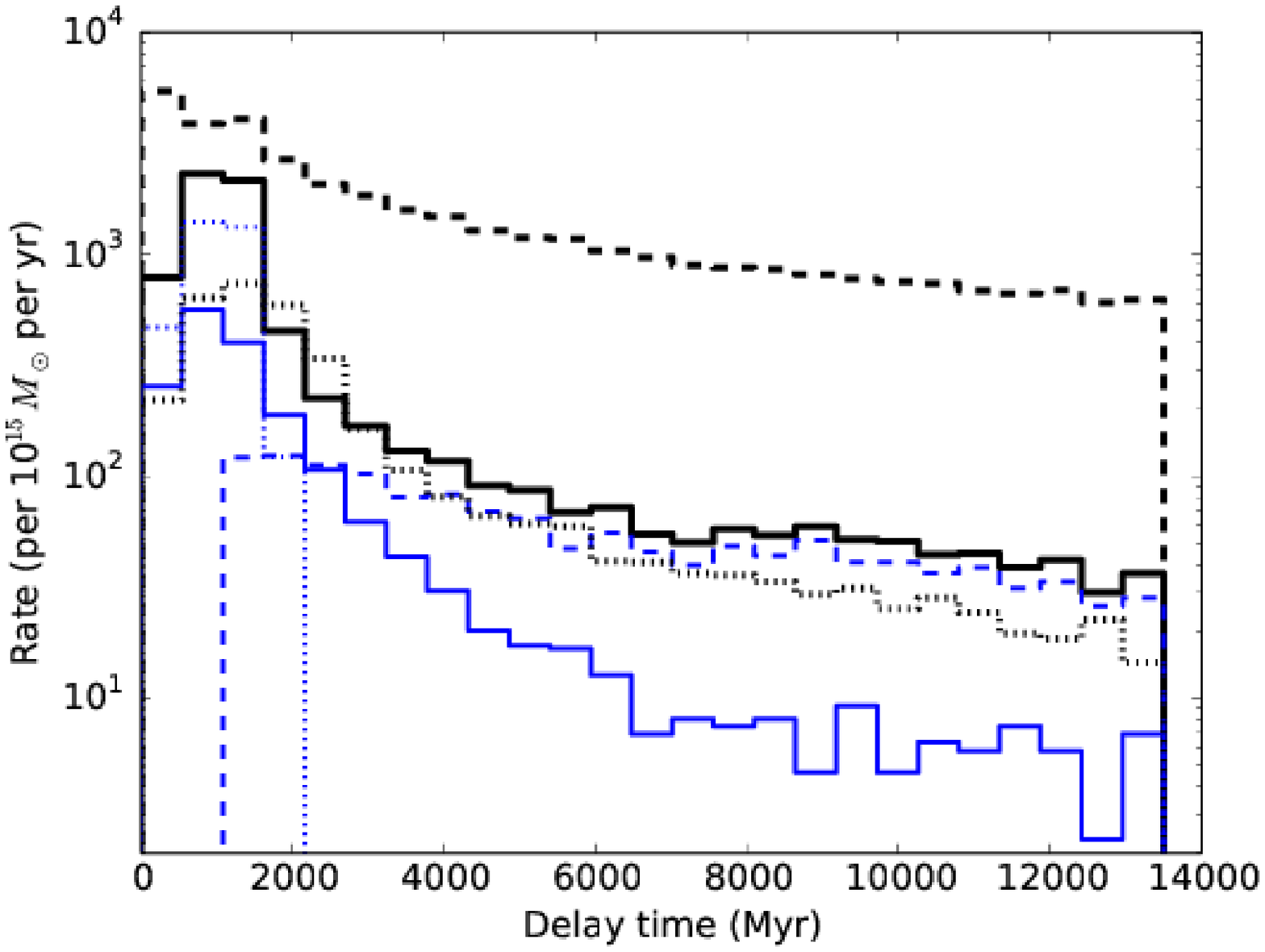}

\caption{\label{fig:rate}The delay time distribution for the formation of
hybrid WDs. Model $\alpha\alpha$ and model $\gamma\alpha$ for the
CE-phase are shown on the left and right, respectively. Dashed (black)
lines depict all type of WDs in interacting binaries, solid (black)
lines depict only hybrid-WDs. Dotted (black) lines show the merger
rate of systems with a hybrid (i.e. their destruction rate), dotted
(thin blue) lines correspond to the main non-degenerate channel (from
more massive stars), while dashed (thin blue) lines depict all hybrids
from the degenerate channel (originating from low-mass stars); solid
(thin blue) show the cases where the secondary is a hybrid, and the
primary is a WD. }
\end{figure*}

We find that hybrid WDs are commonly formed. For a group of stars
with a combined initial mass of 1000${\rm M}_{\odot}$, we expect
3.6-3.9 hybrids to form in a Hubble time. The birthrate of hybrids
as a function of the (delay) time between the zero-age (ZAMS) main-sequence
and the formation of the hybrid is shown in Fig. \ref{fig:rate} (solid
black line). In the first 2 Gyrs, the birthrate is about $10^{-12}$
per yr per solar mass of created stars, and it then decreases to a
few$\times10^{-14}{\rm {yr^{-1}M_{\odot}^{-1}}}$. To put the birthrate
of hybrids in perspective, we also show in Fig. \ref{fig:rate} the
birthrate distribution of white dwarfs of all flavours that are formed
in interacting binaries. One Gyr after star formation, roughly 50-60$\%$
of newly born WDs in compact binaries are hybrids. At later time,
roughly 6$\%$ of the formed-hybrids survive as a hybrid-WDs, since
a significant fraction of these binaries merge and are reduced from
the overall population. 

\subsubsection{Formation channels}

Most hybrids are formed from intermediate mass stars in the first
channel at a time-integrated birthrate of 2.7-3.0$\times10^{-3}$
per ${\rm M}_{\odot}$ of created stars. The masses of the hybrids
span a large range, from a lower limit of $\sim$0.32${\rm M}_{\odot}$
to the upper limit set to 0.63${\rm M_{\odot}}$. The birthrate in
this channel is therefore sensitive to our assumption about the minimum
He fraction of a hybrid. If we are more (less) conservative with our
assumption of what constitutes a hybrid WD, and include hybrids with
He mass fractions only down to $10\%$ ($1\%$), the time-integrated
birthrate from the first channel decreases (increases) to 2.0-2.3$\times10^{-3}$
(3.2-3.4$\times10^{-3}$) per ${\rm M}_{\odot}$ of created stars.
The full range of masses is reached by hybrid progenitors that fill
their Roche lobe on, or before the RGB; if the progenitor loses its
hydrogen envelope on the AGB, the minimum hybrid mass is $\sim$0.52$M_{\odot}$.
The progenitors of the hybrids in this channel have typical ZAMS masses
of 2-4${\rm M}_{\odot}$, and their formation times are short, i.e.
typically several hundred Myrs. This evolutionary channel therefore
gives rise to low-mass WD with masses comparable to those of CO WDs
typically formed only after Gyrs of evolution of single stars. The
formation of hybrid is nevertheless not limited to this short timescale,
but can extend up to several Gyrs for hybrid that form only after
the companion already has become a WD (Fig. \ref{fig:rate}).

The time-integrated birthrate of the second evolutionary channel is
about 8.3-9.4$\times10^{-4}$ per ${\rm M}_{\odot}$ of created stars.
Hybrids from this channel have a mass between $\sim$0.32 and 0.48${\rm M}_{\odot}$.
However, as discussed earlier, such WDs are not likely to be bona-fide
hybrid WDs and produce only minute fractions of CO. Their formation
times are typically long, ranging up to several Gyrs (Fig. \ref{fig:rate}).
Initially, the ZAMS progenitors of the hybrids are 1-2${\rm M}_{\odot}$
stars predominantly in a (circularized) orbit of about 200-500${\rm R_{\odot}}$.
Typically the companion is hydrogen-rich when the hybrid WD is formed.
As can be seen, their overall contribution is small compared to the
first formation channel from more massive progenitors, besides late
times of a few Gyrs.

\subsubsection{Evolution and mergers of hybrid-WDs}

In roughly half ($48-56\%$) of the cases, the stellar components
of the hybrid-WD binaries will eventually merge with one another through
their late evolution. In $25-37\%$ of these mergers the hybrids merge
with hydrogen-rich stellar companions, and in the other systems the
hybrids merge with WD companions. Given their large fractions among
binary WDs, this suggest that a large fraction of all WD mergers involve
a hybrid WD. The merger rate of hybrid white dwarfs is shown in Fig.
\ref{fig:rate} as a function of the merger time (since ZAMS). After
$\sim$4Gyr, a steady state is reached between the formation and destruction
rates of hybrid WDs, such that about there is about 1.5-2 hybrid per
1000${\rm M}_{\odot}$ of created stars (Fig. \ref{fig:lifetime}).
This later-formed WDs, likely contain only a small fraction of CO,
and are effectively He WDs, rather than bona-fide hybrids.

Another likely outcome is that the binary does not change significantly
after the formation of the hybrid, e.g. in cases where the secondary
is a low-mass MS star that does not evolve off the MS in a Hubble
time. This happens for about $\sim20-26\%$ of the hybrids. In $14-31\%$
the binary experiences one or more phases of mass transfer initiated
by the secondary star and forms a double white dwarf system. In about
$2.0-3.9\%$ of systems a cataclysmic variable is formed.

\section{Discussion and summary}

\label{sec:summary}

In this work we have systematically studied the formation and evolution
of hybrid HeCO WDs in binary systems. We studied a wide range of initial
conditions, and explored the distribution of the hybrid-WDs properties
and formation times using binary population synthesis models. Our
findings suggest that hybrid HeCO WDs can form robustly, and give
rise to a significant fraction of all WD binaries with 50-60$\%$
of all young, $<2{\rm Gyr}$ WD being hybrids, (but they become rare
among older populations) In particular, a large fraction of all VLM
WDs, especially young ones, typically considered to be He WDs could
in fact be hybrids. Moreover, the mass range of hybrid WDs can extend
up above to $\sim0.62$ ${\rm M_{\odot}}$, i.e. into the regime typically
considered only for CO WDs. Therefore, observationally hybrid WDs
could be misidentified as CO DB WDs, which could affect mass-radius
interpretations of observations. Similarly, misidentification could
also affect WD chronology estimates, given the difference in the cooling-sequences
for hybrids. The hybrid WDs are composed of significant He abundance,
with He mass fractions in the range $2-25\%$; they reside in an intermediate
mass-radius range between purely He and purely CO WDs. The latter
issue, which is not included in population synthesis studies may slightly
affect the later evolution and interaction of WD-WD binaries, which
can strongly depend on the WD radii. We postpone further exploration
of this issue for future studies. We note that the different structure
and composition of hybrid WDs compared with same mass purely-CO/He
WDs could, in principle be probed using asteroseismology (e.g. \citealt{Win+08}).

Given their prevalence among WD binaries, hybrid WDs may later further
interact with their binary companions. In particular their role in
merger of double degenerate systems (WD-WD, WD-NS, WD-BH) could be
of particular interest. Such hybrids and their mergers could potentially
give rise to explosive thermonuclear events with distinct properties
due to the important role of the He in catalyzing more effective thermonuclear
reactions and detonations (e.g. in the context of Sub-Chandrasekhar
SN explosions \citealt{1986A&A...164L..13B,Woo+86,1990ApJ...361..244L,Ibe+97,Bil+07,2010Natur.465..322P,Wal+11};
WD mergers \citet{2013ApJ...770L...8P}; NS-WD mergers; e.g. \citealt{2016MNRAS.461.1154M};
and WD collisions; e.g. \citealt{2016ApJ...822...19P}). Further studies
of these channels using our newly developed detailed hybrid WD models
will be further explored in forthcoming publications.

\section*{Acknowledgments}
We thank Bill Wolf, Rob Farmer, Ylva Gotberg and Erez Michaely for
stimulating discussions.We acknowledge support from the Israel
Science Foundation I-CORE grant 1829/12. ST acknowledges support from
the Netherlands Research Council NWO (grant VENI {[}\#639.041.645{]}).

\bibliographystyle{mnras}

\hyphenation{Post-Script Sprin-ger}
\begin{thebibliography}{}
\makeatletter
\relax
\def\mn@urlcharsother{\let\do\@makeother \do\$\do\&\do\#\do\^\do\_\do\%\do\~}
\def\mn@doi{\begingroup\mn@urlcharsother \@ifnextchar [ {\mn@doi@}
  {\mn@doi@[]}}
\def\mn@doi@[#1]#2{\def\@tempa{#1}\ifx\@tempa\@empty \href
  {http://dx.doi.org/#2} {doi:#2}\else \href {http://dx.doi.org/#2} {#1}\fi
  \endgroup}
\def\mn@eprint#1#2{\mn@eprint@#1:#2::\@nil}
\def\mn@eprint@arXiv#1{\href {http://arxiv.org/abs/#1} {{\tt arXiv:#1}}}
\def\mn@eprint@dblp#1{\href {http://dblp.uni-trier.de/rec/bibtex/#1.xml}
  {dblp:#1}}
\def\mn@eprint@#1:#2:#3:#4\@nil{\def\@tempa {#1}\def\@tempb {#2}\def\@tempc
  {#3}\ifx \@tempc \@empty \let \@tempc \@tempb \let \@tempb \@tempa \fi \ifx
  \@tempb \@empty \def\@tempb {arXiv}\fi \@ifundefined
  {mn@eprint@\@tempb}{\@tempb:\@tempc}{\expandafter \expandafter \csname
  mn@eprint@\@tempb\endcsname \expandafter{\@tempc}}}

\bibitem[\protect\citeauthoryear{{Abt}}{{Abt}}{1983}]{Abt83}
{Abt} H.~A.,  1983, \mn@doi [\araa] {10.1146/annurev.aa.21.090183.002015},
  \href {http://adsabs.harvard.edu/abs/1983ARA%26A..21..343A} {21, 343}

\bibitem[\protect\citeauthoryear{{Althaus}, {C{\'o}rsico}, {Gautschy}, {Han},
  {Serenelli}  \& {Panei}}{{Althaus} et~al.}{2004}]{Alt+04}
{Althaus} L.~G.,  {C{\'o}rsico} A.~H.,  {Gautschy} A.,  {Han} Z.,  {Serenelli}
  A.~M.,   {Panei} J.~A.,  2004, \mn@doi [\mnras]
  {10.1111/j.1365-2966.2004.07183.x}, \href
  {http://adsabs.harvard.edu/abs/2004MNRAS.347..125A} {347, 125}

\bibitem[\protect\citeauthoryear{{Bildsten}, {Shen}, {Weinberg}  \&
  {Nelemans}}{{Bildsten} et~al.}{2007}]{Bil+07}
{Bildsten} L.,  {Shen} K.~J.,  {Weinberg} N.~N.,   {Nelemans} G.,  2007,
  \mn@doi [\apjl] {10.1086/519489}, \href
  {http://adsabs.harvard.edu/abs/2007ApJ...662L..95B} {662, L95}

\bibitem[\protect\citeauthoryear{{Branch} \& {Nomoto}}{{Branch} \&
  {Nomoto}}{1986}]{1986A&A...164L..13B}
{Branch} D.,  {Nomoto} K.,  1986, \aap, \href
  {http://adsabs.harvard.edu/abs/1986A%26A...164L..13B} {164, L13}

\bibitem[\protect\citeauthoryear{{De Rosa} et~al.,}{{De Rosa}
  et~al.}{2014}]{DeR14}
{De Rosa} R.~J.,  et~al., 2014, \mn@doi [\mnras] {10.1093/mnras/stt1932}, \href
  {http://adsabs.harvard.edu/abs/2014MNRAS.437.1216D} {437, 1216}

\bibitem[\protect\citeauthoryear{{Duch{\^e}ne} \& {Kraus}}{{Duch{\^e}ne} \&
  {Kraus}}{2013}]{Duc13}
{Duch{\^e}ne} G.,  {Kraus} A.,  2013, \mn@doi [\araa]
  {10.1146/annurev-astro-081710-102602}, \href
  {http://adsabs.harvard.edu/abs/2013ARA%26A..51..269D} {51, 269}

\bibitem[\protect\citeauthoryear{{Farmer}, {Fields}, {Petermann}, {Dessart},
  {Cantiello}, {Paxton}  \& {Timmes}}{{Farmer}
  et~al.}{2016}]{2016ApJS..227...22F}
{Farmer} R.,  {Fields} C.~E.,  {Petermann} I.,  {Dessart} L.,  {Cantiello} M.,
  {Paxton} B.,   {Timmes} F.~X.,  2016, \mn@doi [\apjs]
  {10.3847/1538-4365/227/2/22}, \href
  {http://adsabs.harvard.edu/abs/2016ApJS..227...22F} {227, 22}

\bibitem[\protect\citeauthoryear{{Han}, {Podsiadlowski}, {Maxted}, {Marsh}  \&
  {Ivanova}}{{Han} et~al.}{2002}]{Han+02}
{Han} Z.,  {Podsiadlowski} P.,  {Maxted} P.~F.~L.,  {Marsh} T.~R.,   {Ivanova}
  N.,  2002, \mn@doi [\mnras] {10.1046/j.1365-8711.2002.05752.x}, \href
  {http://adsabs.harvard.edu/abs/2002MNRAS.336..449H} {336, 449}

\bibitem[\protect\citeauthoryear{{Heggie}}{{Heggie}}{1975}]{Heg75}
{Heggie} D.~C.,  1975, \mnras, \href
  {http://adsabs.harvard.edu/abs/1975MNRAS.173..729H} {173, 729}

\bibitem[\protect\citeauthoryear{{Hurley}, {Pols}  \& {Tout}}{{Hurley}
  et~al.}{2000}]{Hur00}
{Hurley} J.~R.,  {Pols} O.~R.,   {Tout} C.~A.,  2000, \mn@doi [\mnras]
  {10.1046/j.1365-8711.2000.03426.x}, \href
  {http://adsabs.harvard.edu/abs/2000MNRAS.315..543H} {315, 543}

\bibitem[\protect\citeauthoryear{{Iben} \& {Tutukov}}{{Iben} \&
  {Tutukov}}{1985}]{Ibe85}
{Iben} Jr. I.,  {Tutukov} A.~V.,  1985, \mn@doi [\apjs] {10.1086/191054}, \href
  {http://adsabs.harvard.edu/abs/1985ApJS...58..661I} {58, 661}

\bibitem[\protect\citeauthoryear{{Iben}, {Nomoto}, {Tornambe}  \&
  {Tutukov}}{{Iben} et~al.}{1987}]{Ibe+87}
{Iben} Jr. I.,  {Nomoto} K.,  {Tornambe} A.,   {Tutukov} A.~V.,  1987, \mn@doi
  [\apj] {10.1086/165318}, \href
  {http://adsabs.harvard.edu/abs/1987ApJ...317..717I} {317, 717}

\bibitem[\protect\citeauthoryear{Iben, {Tutukov}  \& {Yungelson}}{Iben
  et~al.}{1997}]{Ibe+97}
Iben I.~J.,  {Tutukov} A.~V.,   {Yungelson} L.~R.,  1997, \mn@doi [\apj]
  {10.1086/303525}, \href {http://adsabs.harvard.edu/abs/1997ApJ...475..291I}
  {475, 291}

\bibitem[\protect\citeauthoryear{{Istrate}, {Marchant}, {Tauris}, {Langer},
  {Stancliffe}  \& {Grassitelli}}{{Istrate} et~al.}{2016}]{Ist+16}
{Istrate} A.~G.,  {Marchant} P.,  {Tauris} T.~M.,  {Langer} N.,  {Stancliffe}
  R.~J.,   {Grassitelli} L.,  2016, \mn@doi [\aap]
  {10.1051/0004-6361/201628874}, \href
  {http://adsabs.harvard.edu/abs/2016A%26A...595A..35I} {595, A35}

\bibitem[\protect\citeauthoryear{{Ivanova} et~al.,}{{Ivanova}
  et~al.}{2013}]{Iva13}
{Ivanova} N.,  et~al., 2013, \mn@doi [\aapr] {10.1007/s00159-013-0059-2}, \href
  {http://adsabs.harvard.edu/abs/2013A%26ARv..21...59I} {21, 59}

\bibitem[\protect\citeauthoryear{{Kroupa}, {Tout}  \& {Gilmore}}{{Kroupa}
  et~al.}{1993}]{Kro93}
{Kroupa} P.,  {Tout} C.~A.,   {Gilmore} G.,  1993, \mnras, \href
  {http://adsabs.harvard.edu/abs/1993MNRAS.262..545K} {262, 545}

\bibitem[\protect\citeauthoryear{{Livne} \& {Glasner}}{{Livne} \&
  {Glasner}}{1990}]{1990ApJ...361..244L}
{Livne} E.,  {Glasner} A.~S.,  1990, \mn@doi [\apj] {10.1086/169189}, \href
  {http://adsabs.harvard.edu/abs/1990ApJ...361..244L} {361, 244}

\bibitem[\protect\citeauthoryear{{Margalit} \& {Metzger}}{{Margalit} \&
  {Metzger}}{2016}]{2016MNRAS.461.1154M}
{Margalit} B.,  {Metzger} B.~D.,  2016, \mn@doi [\mnras]
  {10.1093/mnras/stw1410}, \href
  {http://adsabs.harvard.edu/abs/2016MNRAS.461.1154M} {461, 1154}

\bibitem[\protect\citeauthoryear{{Nelemans}}{{Nelemans}}{2010}]{Nel10}
{Nelemans} G.,  2010, \mn@doi [\apss] {10.1007/s10509-010-0392-0}, \href
  {http://adsabs.harvard.edu/abs/2010Ap%26SS.329...25N} {329, 25}

\bibitem[\protect\citeauthoryear{{Nelemans}, {Verbunt}, {Yungelson}  \&
  {Portegies Zwart}}{{Nelemans} et~al.}{2000}]{Nel+00}
{Nelemans} G.,  {Verbunt} F.,  {Yungelson} L.~R.,   {Portegies Zwart} S.~F.,
  2000, \aap, \href {http://adsabs.harvard.edu/abs/2000A%26A...360.1011N} {360,
  1011}

\bibitem[\protect\citeauthoryear{{Nelemans}, {Portegies Zwart}, {Verbunt}  \&
  {Yungelson}}{{Nelemans} et~al.}{2001}]{Nel+01}
{Nelemans} G.,  {Portegies Zwart} S.~F.,  {Verbunt} F.,   {Yungelson} L.~R.,
  2001, \mn@doi [\aap] {10.1051/0004-6361:20010049}, \href
  {http://adsabs.harvard.edu/abs/2001A%26A...368..939N} {368, 939}

\bibitem[\protect\citeauthoryear{{Pakmor}, {Kromer}, {Taubenberger}  \&
  {Springel}}{{Pakmor} et~al.}{2013}]{2013ApJ...770L...8P}
{Pakmor} R.,  {Kromer} M.,  {Taubenberger} S.,   {Springel} V.,  2013, \mn@doi
  [\apjl] {10.1088/2041-8205/770/1/L8}, \href
  {http://adsabs.harvard.edu/abs/2013ApJ...770L...8P} {770, L8}

\bibitem[\protect\citeauthoryear{{Panei}, {Althaus}  \& {Benvenuto}}{{Panei}
  et~al.}{2000}]{Pan+00}
{Panei} J.~A.,  {Althaus} L.~G.,   {Benvenuto} O.~G.,  2000, \aap, \href
  {http://adsabs.harvard.edu/abs/2000A%26A...353..970P} {353, 970}

\bibitem[\protect\citeauthoryear{{Panei}, {Althaus}, {Chen}  \& {Han}}{{Panei}
  et~al.}{2007}]{Pan+07}
{Panei} J.~A.,  {Althaus} L.~G.,  {Chen} X.,   {Han} Z.,  2007, \mn@doi
  [\mnras] {10.1111/j.1365-2966.2007.12400.x}, \href
  {http://adsabs.harvard.edu/abs/2007MNRAS.382..779P} {382, 779}

\bibitem[\protect\citeauthoryear{{Papish} \& {Perets}}{{Papish} \&
  {Perets}}{2016}]{2016ApJ...822...19P}
{Papish} O.,  {Perets} H.~B.,  2016, \mn@doi [\apj]
  {10.3847/0004-637X/822/1/19}, \href
  {http://adsabs.harvard.edu/abs/2016ApJ...822...19P} {822, 19}

\bibitem[\protect\citeauthoryear{{Paxton}, {Bildsten}, {Dotter}, {Herwig},
  {Lesaffre}  \& {Timmes}}{{Paxton} et~al.}{2011}]{2011ApJS..192....3P}
{Paxton} B.,  {Bildsten} L.,  {Dotter} A.,  {Herwig} F.,  {Lesaffre} P.,
  {Timmes} F.,  2011, \mn@doi [\apjs] {10.1088/0067-0049/192/1/3}, \href
  {http://adsabs.harvard.edu/abs/2011ApJS..192....3P} {192, 3}

\bibitem[\protect\citeauthoryear{{Paxton} et~al.,}{{Paxton}
  et~al.}{2015}]{2015ApJS..220...15P}
{Paxton} B.,  et~al., 2015, \mn@doi [\apjs] {10.1088/0067-0049/220/1/15}, \href
  {http://adsabs.harvard.edu/abs/2015ApJS..220...15P} {220, 15}

\bibitem[\protect\citeauthoryear{{Perets} et~al.,}{{Perets}
  et~al.}{2010}]{2010Natur.465..322P}
{Perets} H.~B.,  et~al., 2010, \mn@doi [\nat] {10.1038/nature09056}, \href
  {http://adsabs.harvard.edu/abs/2010Natur.465..322P} {465, 322}

\bibitem[\protect\citeauthoryear{{Portegies Zwart} \& {Verbunt}}{{Portegies
  Zwart} \& {Verbunt}}{1996}]{Por96}
{Portegies Zwart} S.~F.,  {Verbunt} F.,  1996, \aap, \href
  {http://adsabs.harvard.edu/abs/1996A\%26A...309..179P} {309, 179}

\bibitem[\protect\citeauthoryear{{Prada Moroni} \& {Straniero}}{{Prada Moroni}
  \& {Straniero}}{2009}]{Pra+09}
{Prada Moroni} P.~G.,  {Straniero} O.,  2009, \mn@doi [\aap]
  {10.1051/0004-6361/200912847}, \href
  {http://adsabs.harvard.edu/abs/2009A%26A...507.1575P} {507, 1575}

\bibitem[\protect\citeauthoryear{{Raghavan} et~al.,}{{Raghavan}
  et~al.}{2010}]{Rag10}
{Raghavan} D.,  et~al., 2010, \mn@doi [\apjs] {10.1088/0067-0049/190/1/1},
  \href {http://adsabs.harvard.edu/abs/2010ApJS..190....1R} {190, 1}

\bibitem[\protect\citeauthoryear{{Rebassa-Mansergas}, {Nebot
  G{\'o}mez-Mor{\'a}n}, {Schreiber}, {Girven}  \&
  {G{\"a}nsicke}}{{Rebassa-Mansergas} et~al.}{2011}]{Reb+11}
{Rebassa-Mansergas} A.,  {Nebot G{\'o}mez-Mor{\'a}n} A.,  {Schreiber} M.~R.,
  {Girven} J.,   {G{\"a}nsicke} B.~T.,  2011, \mn@doi [\mnras]
  {10.1111/j.1365-2966.2011.18200.x}, \href
  {http://adsabs.harvard.edu/abs/2011MNRAS.413.1121R} {413, 1121}

\bibitem[\protect\citeauthoryear{{Toonen} \& {Nelemans}}{{Toonen} \&
  {Nelemans}}{2013}]{Too13}
{Toonen} S.,  {Nelemans} G.,  2013, \mn@doi [\aap]
  {10.1051/0004-6361/201321753}, \href
  {http://adsabs.harvard.edu/abs/2013A%26A...557A..87T} {557, A87}

\bibitem[\protect\citeauthoryear{{Toonen}, {Nelemans}  \& {Portegies
  Zwart}}{{Toonen} et~al.}{2012}]{Too12}
{Toonen} S.,  {Nelemans} G.,   {Portegies Zwart} S.,  2012, \mn@doi [\aap]
  {10.1051/0004-6361/201218966}, \href
  {http://adsabs.harvard.edu/abs/2012A%26A...546A..70T} {546, A70}

\bibitem[\protect\citeauthoryear{{Toonen}, {Claeys}, {Mennekens}  \&
  {Ruiter}}{{Toonen} et~al.}{2014}]{Too14}
{Toonen} S.,  {Claeys} J.~S.~W.,  {Mennekens} N.,   {Ruiter} A.~J.,  2014,
  \mn@doi [\aap] {10.1051/0004-6361/201321576}, \href
  {http://adsabs.harvard.edu/abs/2014A%26A...562A..14T} {562, A14}

\bibitem[\protect\citeauthoryear{{Toonen}, {Hollands}, {G{\"a}nsicke}  \&
  {Boekholt}}{{Toonen} et~al.}{2017}]{Too17}
{Toonen} S.,  {Hollands} M.,  {G{\"a}nsicke} B.~T.,   {Boekholt} T.,  2017,
  \mn@doi [\aap] {10.1051/0004-6361/201629978}, \href
  {http://adsabs.harvard.edu/abs/2017A%26A...602A..16T} {602, A16}

\bibitem[\protect\citeauthoryear{{Tutukov} \& {Yungelson}}{{Tutukov} \&
  {Yungelson}}{1992}]{Tut+92}
{Tutukov} A.~V.,  {Yungelson} L.~R.,  1992, \sovast, \href
  {http://adsabs.harvard.edu/abs/1992SvA....36..266T} {36, 266}

\bibitem[\protect\citeauthoryear{{Waldman}, {Sauer}, {Livne}, {Perets},
  {Glasner}, {Mazzali}, {Truran}  \& {Gal-Yam}}{{Waldman}
  et~al.}{2011}]{Wal+11}
{Waldman} R.,  {Sauer} D.,  {Livne} E.,  {Perets} H.,  {Glasner} A.,  {Mazzali}
  P.,  {Truran} J.~W.,   {Gal-Yam} A.,  2011, \mn@doi [\apj]
  {10.1088/0004-637X/738/1/21}, \href
  {http://adsabs.harvard.edu/abs/2011ApJ...738...21W} {738, 21}

\bibitem[\protect\citeauthoryear{{Winget} \& {Kepler}}{{Winget} \&
  {Kepler}}{2008}]{Win+08}
{Winget} D.~E.,  {Kepler} S.~O.,  2008, \mn@doi [\araa]
  {10.1146/annurev.astro.46.060407.145250}, \href
  {http://adsabs.harvard.edu/abs/2008ARA%26A..46..157W} {46, 157}

\bibitem[\protect\citeauthoryear{{Woosley}, {Taam}  \& {Weaver}}{{Woosley}
  et~al.}{1986}]{Woo+86}
{Woosley} S.~E.,  {Taam} R.~E.,   {Weaver} T.~A.,  1986, \mn@doi [\apj]
  {10.1086/163926}, \href {http://adsabs.harvard.edu/abs/1986ApJ...301..601W}
  {301, 601}

\bibitem[\protect\citeauthoryear{{Zhang}, {Hall}, {Jeffery}  \& {Bi}}{{Zhang}
  et~al.}{2018}]{Zha+18}
{Zhang} X.,  {Hall} P.~D.,  {Jeffery} C.~S.,   {Bi} S.,  2018, \mn@doi [\mnras]
  {10.1093/mnras/stx2747}, \href
  {http://adsabs.harvard.edu/abs/2018MNRAS.474..427Z} {474, 427}

\makeatother
\end{thebibliography}

\bsp	
\label{lastpage}
\end{document}